\documentclass[11pt]{article}
\usepackage{graphicx}
\usepackage{tabularx}

\usepackage{wrapfig}
\usepackage{float}
\usepackage{amsmath, amssymb, amsthm}
 
\usepackage{setspace}
\usepackage[usenames,dvipsnames,svgnames,table]{xcolor}
\definecolor{lightgray}{gray}{0.9}
\definecolor{gray}  {RGB}{231,231,231} 
\definecolor{RoyalBlue}  {RGB}{47,0,255} 
\definecolor{RoyalPurple}  {RGB}{197,0,130} 
\definecolor{RoyalGreen}  {RGB}{217,231,201}
\definecolor{lightblue}  {RGB}{196,213,245}
\definecolor{lightviolet}  {RGB}{199,209,214}

\usepackage{array}
\usepackage{listings}
\usepackage{fullpage}

\usepackage{fancyhdr,lipsum}
\usepackage{caption}
\usepackage[perpage, flushmargin, symbol*]{footmisc}
  \let\oldfootnote\footnote
   \renewcommand\footnote[1]{%
    \oldfootnote{\hspace{1 mm}#1}}

\usepackage[pdftex, plainpages = false, pdfpagelabels, 
            pdfpagelayout = OneColumn,
            bookmarks,
            bookmarksopen = true,
            bookmarksnumbered = true,
            breaklinks = true,
            linktocpage,
            pagebackref = false,
            colorlinks = true,
            linkcolor = black,
            urlcolor  = black,
            citecolor = black,
            anchorcolor = green,
            hyperindex = true,
            hyperfigures
            ]{hyperref} 

\usepackage[natbibapa]{apacite}

\usepackage{tikzrput}
\usepackage[object=vectorian]{pgfornament}
\RequirePackage{tikz}

\begin{document}

\thispagestyle{plain}
\begin{flushleft}
\textbf{\Large An Empirical Unravelling of Lord's Paradox}\\
\vspace{0.5cm}
\textbf{ZhiMin Xiao\textsuperscript{a,}\footnote{\small Corresponding author. \newline E-mail: zhimin.xiao@durham.ac.uk}, Steve Higgins\textsuperscript{b}, Adetayo Kasim\textsuperscript{c}}
\vspace{1 em}

\small \textsf{\textsuperscript{a,b} School of Education, Durham University, Durham, DH1 1TA, UK}\newline
\small \textsf{\textsuperscript{c} Wolfson Research Institute, Queen's Campus, Durham University, Stockton-on-Tees, TS17 6BH, UK}

\end{flushleft}

\subsection*{Abstract}
Lord's Paradox occurs when a continuous covariate is statistically controlled for and the relationship between a continuous outcome and group status indicator changes in both magnitude and direction. This phenomenon poses a challenge to the notion of evidence-based policy, where data are supposed to be self-evident. We examined 50 effect size estimates from 34 large-scale educational interventions, and found that impact estimates are affected in magnitude, with or without reversal in sign, when there is substantial baseline imbalance. We also demonstrated that multilevel modelling can ameliorate the divergence in sign and/or magnitude of effect estimation, which, together with project specific knowledge, promises to help those who are presented with conflicting or confusing evidence in decision making.

\subsection*{Keywords} 
Evaluation; Evidence-Based Policy; Lord's Paradox; Multilevel Modelling; RCT

\section*{The origin of Lord's Paradox}

Frederic M. Lord of the Education Testing Service presented a hypothetical situation in 1967 where a dietician was interested in knowing the effect of food provided in a university's dining halls on the weight of individual male and female students, which was measured at the beginning and end of an academic year. When the data became available, two statisticians were called upon to independently investigate if there was any dietary effect on body weight for the two sexes. Using gain score analysis or a paired $t$-test \citep[p.~429]{tu2010}, the first statistician found that neither the average weight nor the frequency distribution of weight for males or females changed during the academic year, indicating no evidence of dietary effect on weight gain for either males or females. Conducting an analysis of covariance (ANCOVA) on the same data, the second statistician asserted that wherever the two subgroups started with similar initial weight, males gained much more weight than females did throughout the academic year \citep{lord1967}.

The above statistical phenomenon, called Lord's Paradox, is related to, but less well-known than Simpson's Paradox, which reveals one trend when a whole body of data is analysed, but quite the opposite one when the same data are split into subgroups \citep[p.~268]{ONeil2016}. The primary difference between the two paradoxes, however, lies in the types of variables considered, the former occurs in ANCOVA where the outcome variable and baseline measurement adjusted for are both continuous, whereas the latter arises when a categorical independent variable is controlled for \citep{tu2010}. They are both examples of ``the reversal paradox,'' where the association between two variables, one independent and the other dependent, reverses, diminishes, or enhances when a third covariate is statistically controlled for \citep{Tu2008}.

\section*{Why is it important to understand and resolve the Paradox?}

Understanding when the reversal paradox arises is important for decision making, where data are sometimes supposed to be self-evident. As illustrated in Lord's dining hall example, the conclusions both statisticians made were ``visibly correct'' \citep[p.~305]{lord1967} and backed up by the evidence they had on hand. And yet, when different analytical approaches were applied to the same data, the estimate of effect changed in both direction and magnitude. This phenomenon poses even greater challenges to evidence-based practice and decision making than in the case of covariate selection, which can potentially produce any desired evidence should a sufficient number of variables be explored in a given study \citep{Simmons2011}. 

For research funders, such as the Education Endowment Foundation (EEF) in England, comparable estimates of impact from across a large number of projects are essential in answering effectiveness questions, such as which interventions have worked and might be made to work and for whom in English schools. If the inferences remain problematic due to the variety of research designs and the diversity of analytical approaches, such as the ones introduced in Lord's hypothetical case, decisions on the allocation of research resources could be unnecessarily costly, let alone any ethical implications the interventions might have on the teachers and pupils of participating schools. As the independent grant-making charity intends to award as much as \pounds 200 million by 2025, so that pupils from all backgrounds in England can reach their full potential and, the association between socioeconomic background and academic attainment can be weakened, explorations of potential methods that promise to minimise the discrepancy in analytical outcomes become ever more crucial. This piece of research aims to find out if and when the reversal paradox occurs in EEF-funded studies and seeks to find a solution that has the potential to reduce the variation in effect size estimates resulting from similar analytical approaches to those seen in Lord's case of dietary effect. 

The EEF has taken steps to maximise the objectivity and validity of the interventions they fund. One example is to commission independent evaluations of those interventions. But rigorous evaluations have their limitations, partly because they are independent, which means there is a trade-off between the most appropriate evaluation strategy for a given intervention and the comparability of results from across the studies that are designed, implemented, and analysed in diverse ways. The comparability issue can be partially solved when outcomes, covariates, and analytical models are pre-specified, after taking into consideration specific research designs, the quality as well as the structure of the data \citep{zmxaksh2016a}. But in situations where straightforward answers are not easy to come by, as in Lord's Paradox, further explorations of the reversal paradox are necessary when such decisions must be made. 

Most extant literature seems to suggest that Lord's Paradox only occurs in observational studies \citep{Knapp2009, lord1967, lord1969, senn2006, tu2010, Tu2008, Breukelen2006, VanBreukelen2013, Wainer2004}. In concluding the original hypothetical case paper, Lord (1967) noted: ``with the data usually available for such studies, there simply is no logical or statistical procedure that can be counted on to make proper allowances for uncontrolled pre-existing differences between groups'' (p. 305). This suggests that neither ANCOVA nor gain score analysis would help to solve the puzzle. While ANCOVA seems a logical approach amongst many behavioural scientists and where baseline balance cannot be achieved, it uses within-group regression lines to make the adjustment, which can be useful, as Lord (1969) argued later, if real life problems allow us to discard certain extreme units of analysis within those groups. Otherwise, ``no reason can be advanced why the within-groups regression line should give the proper adjustment'' (p. 337). In other words, we cannot explain why we should statistically coerce the units of analysis from different comparison groups to have the same baseline measure and then compare their expected means \citep[p.~430]{tu2010}. 

Regarding gain score analysis, Lord (1969) also mentioned that many researchers would ``shrug off'' the Paradox by subtracting the baseline measure from final outcome. But the difference in change approach is equally ``illusory'' (p. 336), because ``the difference $y - x$ is a quantity of no interest except when $y$ and $x$ both measure the same dimension'' (p. 336) and when the difference between scores is as reliable as the pre- and post-test results themselves \citep[p.~161]{Dimitrov2003}. In education, the level of difficulty in a test usually adds another layer of complexity to the problem, as equal gains in test scores do not always represent equal changes in capability. If a test is relatively easy, a gain score approach might incorrectly indicate more progress for pupils of lower attainment. Conversely, if a test is relatively difficult, the estimate from gain score analysis might falsely indicate more progress for pupils of higher attainment \citep[pp.~162-163]{Dimitrov2003}. 

To rule out the possibility that the observed difference in outcome is not due to pre-existing differences at baseline, Lord implicitly suggested that one should randomly assign units of interest (e.g. students or plants) to experimental conditions (1969, p. 336), because randomisation allows us to see how the groups would have compared should there be no difference at baseline. The question asked of the two statisticians in the dietary case is a causal one without a comparison group, but the effect of any intervention is a relative concept \citep[see also ][p.~121]{Wainer2004}, where the treatment group must be compared with a counterfactual, an equivalent control group that received an alternative or no treatment. As Holland (2005) pointed out, there was no control group in Lord's hypothetical scenario, where all students ate in the dining halls, which renders the case a pre-post single group design. 

So, an appropriate answer to the question depends on the assumptions the two statisticians made about their hypothetical control groups. The first one assumed that the ``control group'' weight (i.e., the final weight of each student in June had the student not eaten in the dining hall throughout the academic year) would be the same as the student's initial weight at the beginning of the academic year. While the second one also assumed that final weight is a linear function of initial weight, but the analytical approach requires the two within-group regression lines to be parallel to each other and differ only in their intercepts. Unfortunately, none of the above-mentioned assumptions was testable \citep{Wainer2004}. The two statisticians could be right or wrong, depending on which assumption one is prepared to make and accept. In Lord's words: ``There are as many different explanations as there are explainers'' (1967, p. 305). To further explore the phenomenon, let's turn to EEF studies where there are at least two comparison groups. 

\section*{Methods}
\subsection*{Description of EEF research designs}

This study uses data from 34 completed EEF projects, which vary in design and quality of implementation, as shown in Table \ref{fullname}. The EEF has funded over 100 projects, but only the data from the first 34 had been made available in an archive at the beginning of this analysis. We believe that the samples are large enough for us to properly explore the phenomenon, as the results will show later. The 34 projects represent six broad research designs, which are \textit{srt}, \textit{mst}, \textit{crt}, \textit{action}, \textit{quasi}, and \textit{rdd} in the \textit{design} column of Table \ref{fullname}. The first three are experimental designs, referring to simple, multi-site, and cluster randomised trials respectively. In \textit{srt}, individual students are randomly allocated to intervention or control. This design is relatively simple and easy to understand, but often generates imbalanced groups in terms of sample size and baseline imbalance even in large studies \citep[pp.~30-31]{Torgerson2008}. In \textit{mst}, randomisation still occurs at student level, but it takes place in multiple schools, and the blocking is intentional, meaning all participating schools have sufficiently large numbers of students in either intervention or control group. Most early EEF projects can be classified as \textit{mst}. Nevertheless, it can sometimes be difficult to differentiate \textit{srt} from \textit{mst} in larger studies, where pupils happen to come from many schools and some unintentionally perfect blocking effect is achieved \citep[p.~4]{zmxaksh2016a}. In \textit{crt}, clusters, such as schools, classes, or year groups, are randomly assigned to treatment arms. While the approach can effectively minimise contamination that could be a problem in \textit{srt} and \textit{mst}, observations within the same clusters are often correlated, which violates the independence assumption of standard statistical methods \citep[p.~5]{hayes2009}. However, the violation justifies the adoption of multilevel modelling in this study.

The last three of the six research designs listed in Table \ref{fullname} are action research, quasi-experiments, and regression discontinuity design (\textit{rdd}), which are all observational studies but involve comparison groups. It is worth mentioning that we are not the evaluation teams who independently designed the 34 projects. We have access to the data and accepted the processes that generated them. Therefore, the definitions we provide here are descriptive of the approaches used rather than driven by theory or literature. This is more apparent in the three non-experimental designs, where, due to the nature of the studies being a pilot and/or having substantial imbalance at baseline, threats to internal validity are greater than those in the first three categories, where the design, implementation, and reporting usually followed CONSORT standards \citep{Altman2012}. Now let's turn to the few studies that were classified as observational or quasi-experimental designs.

Effective Feedback is an action research project, where teachers were encouraged to utilise the research evidence on feedback \citep[e.g.][]{Hattie2007} in their teaching. It was a one-year pilot study, where both teachers' and pupils' understanding of the intervention were constantly monitored and the intervention adjusted accordingly at the end of each action research cycle. The study took place in nine intervention and five comparison schools in London, but randomisation did not occur at any level or stage \citep{Siddiqui2014}. The second observational design in Table \ref{fullname} is Tutor Trust Primary and Secondary, which aimed to provide affordable small group or one-to-one tuition to disadvantaged pupils in challenging communities. Although the study was set up to test the intervention under realistic conditions in many schools, the quasi-experimental design through matching was unable to establish a comparison group who were similar in demographic and socio-economic characteristics due to attrition and ``marked differences'' in prior attainment \citep[p.~4]{tt2015}. SHINE in Secondaries is the last quasi-experimental design included in the study. The pilot project aimed to improve pupil attainment by focusing on literacy and numeracy and revisiting areas where pupils struggled most through a creative curriculum in four schools. The evaluation team employed an \textit{rdd} to assess the effect of the Saturday programme, where pupils were assigned to intervention if they fell below the first cut point on pre-test, to control if they stood above a second cut point. Those who were between the two cut points were randomly assigned to intervention or control. This approach guaranteed all low-attaining pupils had the opportunity to participate in the intervention \citep[p.~4]{shine2016}. However, \textit{rdd} is less powerful than a conventional experimental design and it requires all pupils to remain in the group to which they were originally assigned. Otherwise, the inference and power may be further compromised \citep[p.~11]{shine2016}.

\subsection*{Four analytical models to study the Paradox}

The research designs explained above were therefore beyond our control and should be differentiated from the methods we employ in this study to explore the Paradox in question. In effect, we have simply re-analysed the same dataset for each of the 34 projects four times using four different analytical models. The first two compare the effect estimates from difference-in-means of post-test and gain scores, and the last two investigate if there is any difference in estimates from two multilevel models, one having post-test as the outcome and controlling for pre-test and group status, namely, post-ANCOVA; and the other having gain score as the outcome but including no other covariate than group status, which is gain-ANOVA or ``ANOVA of change'' \citep{Breukelen2006}. Both multilevel models treat schools as the only random variable.

\subsubsection*{Difference-in-means}

In an ideal experimental study where a large number of individuals are randomly allocated to treatment or control and the implementation is successful with equal sample sizes per intervention arm and very little missing data, a difference-in-means of post-test alone should be able to estimate a similar average treatment effect to that of ANCOVA \citep[p.~1123]{Vickers2001}, although the latter is often considered to have more power than the former in detecting an effect \citep{Petscher2011, tu2010, Tu2008a, Breukelen2006, Vickers2001}. Nonetheless, real world experimental designs are usually more complex than the one described above, as shown in the \textit{design} column in Table \ref{fullname} and the columns that involve sample sizes in Table \ref{stats}. In fact, only three (rows 12, 13, and 14 in Table \ref{fullname}) out of the 50 outcomes are classified as \textit{srt}, which is closest to the aforementioned design. This fact suggests that a simple difference-in-means of post-test scores alone would be problematic, or put in another way: ``the simplicity is deceiving'' \citep[p.~2]{Knapp2009}, as the analysis strategy does not fully take the research design into account as any variation at baseline would come from within and between groups \citep[p.~5]{Tu2008}. 

In fact, rarely would anybody compare the difference in post-test alone without examining how successful the randomisation actually was in achieving balance at baseline. In some EEF evaluations, a simple difference-in-means of post-test scores does appear, but only when pupils in treatment and control groups are similar in average pre-test achievement. When the two groups differ substantially at baseline, a difference-in-means of gains is sometimes used instead. We therefore introduce this set of comparisons because it has practical implications for the EEF, as the two methods do occur in projects that are selected for this study (for instance, outcomes labelled as \texttt{ffe}, \texttt{ffm}, \texttt{sor}, \texttt{efr}, \texttt{efw}, and \texttt{efm} in Table \ref{fullname}) and beyond (more completed EEF projects evaluated using difference-in-means of either post-test or gain scores joined our data archive after the submission of this paper). 

Evaluation teams often compare the difference in average performance at pre-test between intervention and control groups to check if balance on the key variable was indeed achieved. To standardise this, we converted the difference-in-means of pre-test scores between intervention and control groups into Hedges’ $g$ so that the level of imbalance on pre-test was as comparable as possible across all the studies selected for this exploration. The equations used to compute the imbalance metric are the same as Equations 1 through 4 in \citet[p.~5]{zmxaksh2016a}. The only difference is that this time the quantity of interest is pre-test rather than post-test.

While groups in most of the EEF studies are balanced at baseline, as would be expected with randomisation, they do occasionally differ considerably. As shown in the \textit{pret.imb} column of Table \ref{stats}, the bottom eight out of those 50 rows have absolute values greater than 0.2, which means the students in intervention and control groups differ substantially in baseline attainment, with pupils in treated groups being at least three months \citep[see][for how the month-in-progress is calculated]{toolkit2015} ahead of those in control groups when the value is positive (or the other way around when the value is negative). In fact, four out of the eight having standardised differences at baseline greater than 0.3, and those outcomes are \texttt{fs}, \texttt{cmtm}, \texttt{ttsm}, and \texttt{shine}. In all of the four cases, pupils in the intervention groups performed a lot lower than those in the control groups, hence negative in sign.

However, imbalance is not the only issue worth consideration. As noted earlier, when initial and final measures are not comparable and when there are measurement errors, a gain score analysis involving the transformation of pre- and post-tests into $z$-scores can be problematic. Switching from difference-in-means of post-tests to that of gains, which is effectively equivalent to predicting post-test with an ANCOVA that controls for pre-test \citep{Pascarella2003}, is supposed to adjust for imbalance at baseline, although some argue that it does not because of regression to the mean \citep[see][p.~1123]{Vickers2001} or the fact that pre-test scores are usually negatively correlated with gains \citep{Knapp2009}. If it is true, then there is not much space for more sophisticated models in the analysis of experimental data, regardless of the design that has generated the data and the quality and structure of the data. 

\subsubsection*{Multilevel modelling}

In Lord's original dietary case, there was only one university. But if we suppose several universities are interested in the study and their students are from geographically and ethnically diverse backgrounds in the United States, the statisticians, particularly the first one, might have to think twice before they proceed with their preferred analytical strategy, as those diners might have very different average weights at the beginning of the academic year. In that case, even if initial average weights between the groups are similar, the big variation from dining hall to dining hall might render either gain score analysis or ANCOVA insufficient to capture some key features such as variation in the data. In addition, if there were more students in some of the dining halls and fewer in others, the different group sizes would also affect the analysis. That is why we also include, for better comparison, a second set of results from the two multilevel models, to which we now provide a detailed introduction in the paragraphs that follow.

While it is not uncommon that gain or difference scores are used as the outcome for ANCOVA \citep{Pascarella2003, Pike2004}, the dependent variable of the ANCOVA in this study is primary post-test results for all the projects considered. Although the two approaches are mathematically equivalent, as shown in the two equations below: 

\begin{equation*}
post_{i} - pret_{i}=\beta_0 + \beta_1T_{i} + \epsilon_{i},
\end{equation*}

\noindent where $post_{i} - pret_{i}$ represents gains from pre-test to post-test, $\beta_0$ is the intercept, $\beta_1$ is average treatment effect, and $T_{i}$ is treatment indicator. $\epsilon_{i} \sim N(0,\sigma^2)$ refers to the residual component of the model. If we add $pret_{i}$ to both sides of the above equation, which is a $t$-test in a regression framework \citep[p.~292]{Tu2008a}, we get:

\begin{equation*}
post_{i}=\beta_0 + \beta_1T_{i} + \beta_2pret_{i} + \epsilon_{i},
\end{equation*}
\noindent which is equivalent to the first one \citep[for a similar argument where gain-ANCOVA also controls for baseline measure, see][pp.~1-2]{Eriksson2014}, although the implicit assumptions underpinning the two equations are different \citep[p.~293]{Tu2008a}. That said, we still prefer post-ANCOVA to gain-ANCOVA for the following reasons.

First, as mentioned earlier, not all projects have equivalent pre- and post-test measures. When baseline and final measures are on different scales, transforming them into comparable scales so that the difference can be calculated may, in our view, impose a rather different distribution to the data, let alone that there are usually multiple ways to transform the data \citep[p.~356]{McElreath2016}. That is to say, arbitrary data transformations might impede the comparability of effect estimates from across the studies when the transformation alters the distributions under concern \citep{zmxaksh2016a}. Second, using gain score as the outcome variable, with or without baseline measurement statistically controlled for, answers a slightly different question \citep[p.~352]{Pike2004}, which concerns the difference in change of attainment between intervention and control groups \citep[p.~292]{Tu2008a}. When the outcome is post-test and pre-test a covariate, the question would relate to the difference in post-test between the two groups who had similar scores on pre-test \citep[p.~2]{Knapp2009}. It was shown elsewhere that balance at baseline, ``not even in expectation'' \citep[p.~4334]{senn2006}, is not a necessary condition for post-ANCOVA to be unbiased. Most EEF-funded projects are randomised experimental designs, which assume the intervention groups would, on average, be similar in all aspects, including baseline measurement \citep[p.~5]{Tu2008}. In fact, the first 30 pre-test imbalance scores in Table \ref{stats} are no greater than 0.1, only eight out of the 50 outcomes have a score greater than 0.2 in magnitude. For these reasons, we therefore consider post-test results the most appropriate and comparable outcome of interest for the purpose of this research (indeed, most evaluation teams used post-test as the outcome variable). 

Having justified the choice of post-test scores as the outcome variable, we should also explain why we also use multilevel models. Most data in education is nested. Pupils are taught in classes in schools. Many EEF trials are classified as \textit{crt}, which takes account of this clustering at school level. Other analytical techniques which do not take account of this will produce similar estimates when the schools are very similar and the sample evenly drawn from the schools. But it is often not the case. We have argued elsewhere that effect size estimates derived from a multilevel model's total variance should be used because they are less prone to false positives than models that do not take account of the data structure and research design \citep{zmxaksh2016a}. 

As we allow schools to differ in their average attainment by introducing a random intercept at school level in the two multilevel models \citep[for the mathematical equations that underpin the models, please see][pp.~5-7]{zmxaksh2016a}, they thus perform a sort of partial pooling \citep{Gelman2012}, where schools with larger sample sizes ``lend'' information to those with fewer pupils. The pooling effect is analogous to the value of ``remembering,'' which helps us improve accuracy in prediction, as the model makes it possible to simultaneously identify the features of individual schools and learn about all the schools \citep[pp.~355-356]{McElreath2016}. 

Also, partial pooling of multilevel modelling in the second set of comparison contrasts with complete or no pooling in the difference-in-means models of the first set, where information about one school told us nothing about others, and effects within individual schools were assumed to be homogeneous with only sampling errors, regardless of sample sizes in the schools \citep[p.~364]{McElreath2016}. Those assumptions are often untenable, for instance, equal sample sizes for all the schools involved in a given study are rare. Furthermore, there are at least two risks in averaging instead of modelling. First, when there is imbalance in sample size across the schools, any inference would be unfairly dominated by those with larger samples. Second, averaging almost always reduces variation, which is more likely to manufacture ``false confidence'' \citep[p.~356]{McElreath2016}. Despite the caveats associated with the simpler models, they have been used to evaluate some EEF projects. For this reason, we consider it appropriate to compare the results from the difference-in-means approach to those from the two multilevel models. Given the data we have access to and the models we have defined, we now turn to the results. 

\section*{Results}

It is important to note that headline effect estimates in EEF evaluation reports, which are available for the public to download on EEF's website, should not be compared against the estimates from the models employed in this study, as evaluation models are diverse, serve very different purposes, and use a variety of covariates \citep[p.~3]{zmxaksh2016a}. To be exact, the evaluation models aimed to identify the impact on pupil's learning due to the intervention. In this study, we aimed to explore Lord's Paradox so applied identical analytical models. Also, in order to calculate gain scores for all the projects included, we converted some of the pre- and post-test scores, when they were not on the same scale, into $z$-scores for comparability.

\begin{figure}[h]
\begin{center}
\includegraphics[scale=.7]{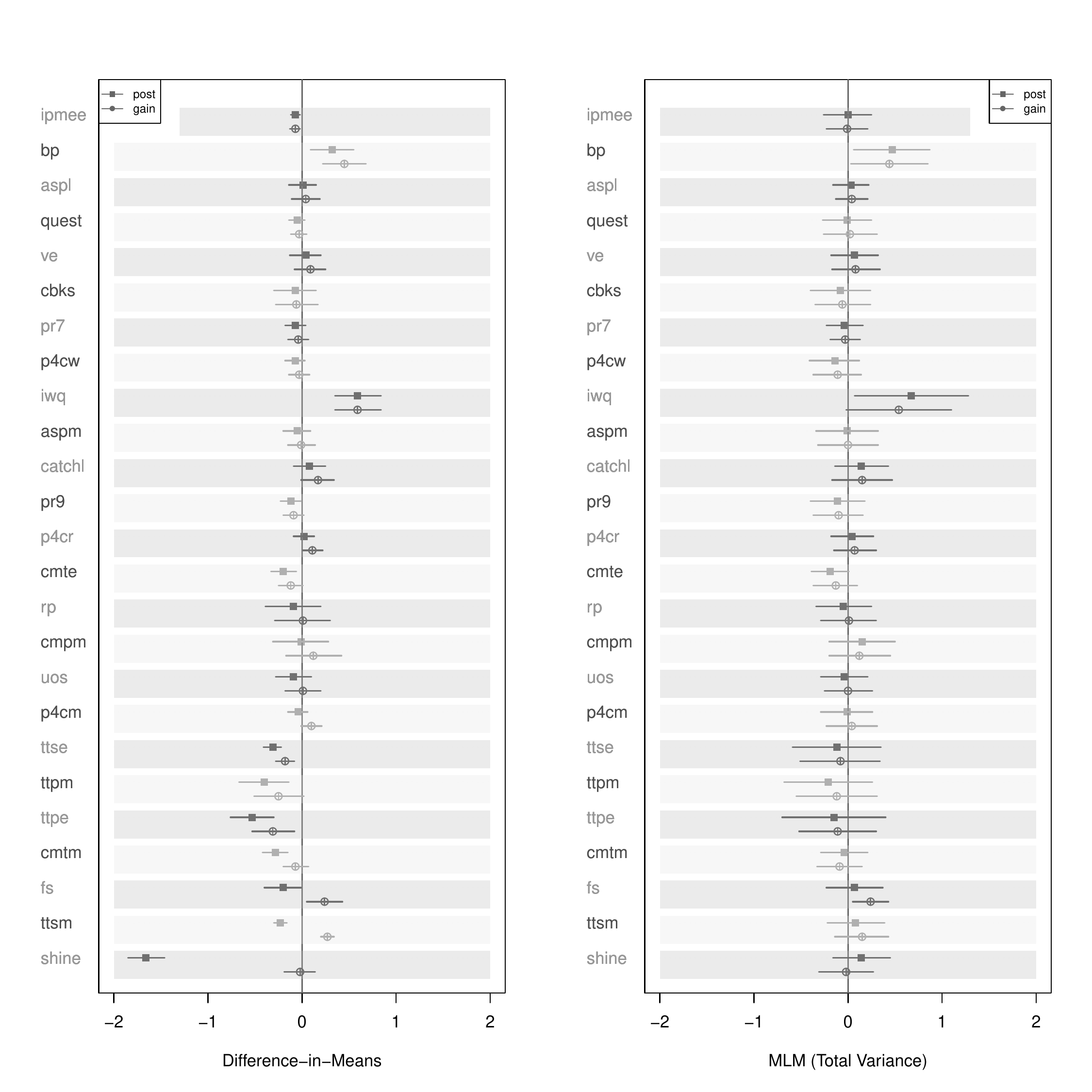}
\caption{{\bf Comparison of effect size estimates for outcomes that have negative pre-test imbalance values.}
For each outcome, there are two sets of effect size estimates. The first set, \textit{post}, represents models using post-test as the outcome variable. The second set, \textit{gain}, has gain score as the outcome variable.}
\label{Lord1to25}
\end{center}
\end{figure}

\begin{figure}[h]
\begin{center}
\includegraphics[scale=.7]{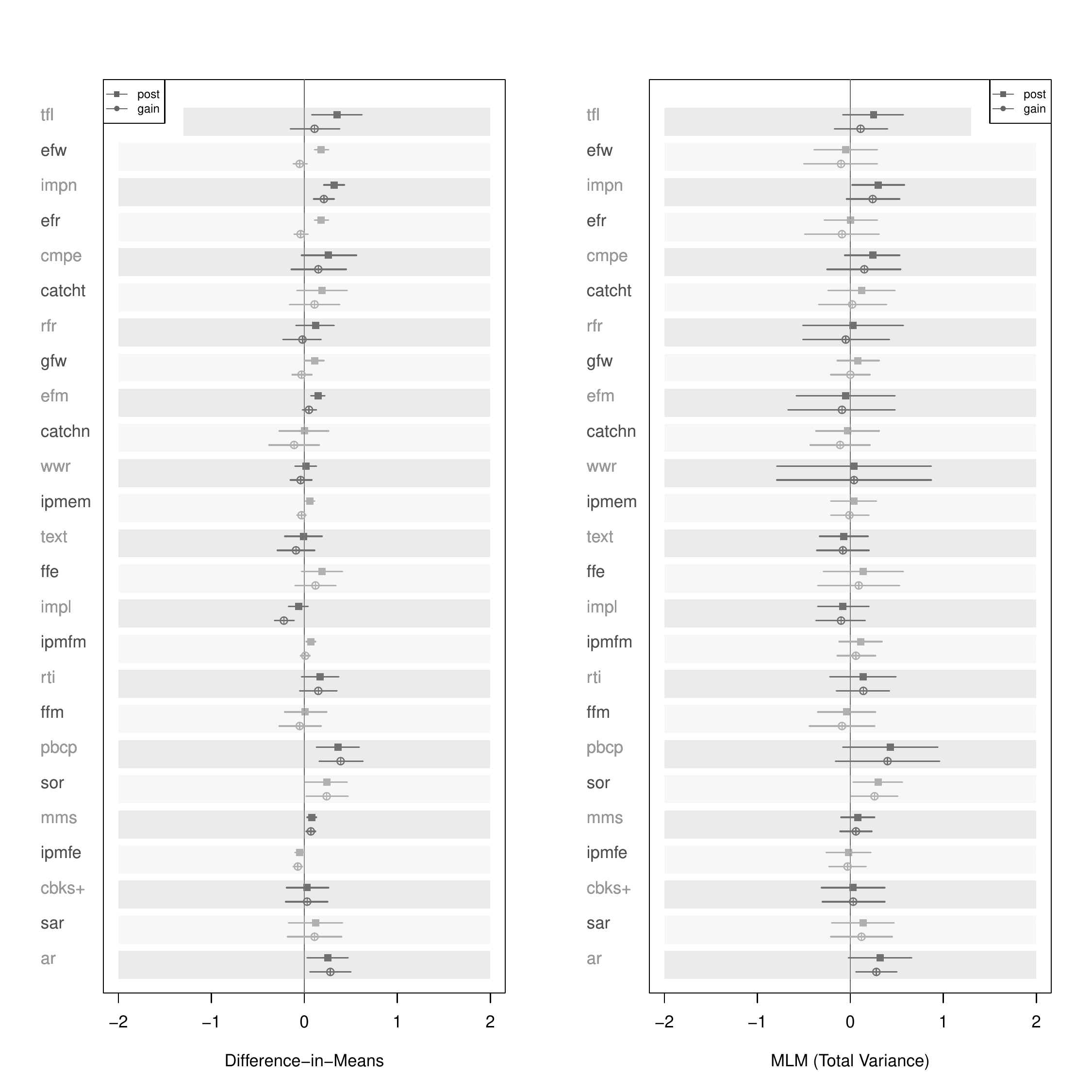}
\caption{{\bf Comparison of effect size estimates for outcomes that have non-negative pre-test imbalance values.}}
\label{Lord26to50}
\end{center}
\end{figure}

Once we computed the results from the four analytical models for all the projects included, we plotted them into Figures \ref{Lord1to25} and \ref{Lord26to50}, where the estimates from the two difference-in-means of post-test and gain scores are listed in the left panels, and those from the two multilevel models, namely, post-ANCOVA and gain-ANOVA, are in the right panels. Let's first focus on the left panels of the two figures. The 50 outcomes are organised such that all estimates in Figure \ref{Lord1to25} have negative pre-test imbalance values, and all those in Figure \ref{Lord26to50} have non-negative ones. Pre-test imbalance in the study is measured as the standardised difference-in-means of pre-test scores between intervention and control groups. The values are all Hedges’ $g$, which corrects for the bias in estimates when sample sizes are small. Since the difference is calculated as the mean of intervention group on pre-test minus that of control group, a negative value indicates that the intervention group performed less well, on average, at baseline than did the control group. Nevertheless, the values can also be zero or positive. The data behind the ordering are presented in Table \ref{efs}, where the \textit{pret.imb} column lists the above-mentioned imbalance measures from the most negative to the most positive, meaning the most imbalanced studies at baseline appear at the two ends of the list, those in the middle have the most similar intervention and control groups in terms of performance on pre-test.

The project \texttt{shine} in Figure \ref{Lord1to25} has the largest negative pre-test imbalance score at $-2.64$, and \texttt{tfl} in Figure \ref{Lord26to50} has the largest positive imbalance at $0.25$. However, for the convenience of reading and referencing, outcomes are presented, from top to bottom, in either ascending (Figure \ref{Lord1to25}) or descending (Figure \ref{Lord26to50}) order of their corresponding pre-test imbalance scores. For instance, in Figure \ref{Lord1to25}, \texttt{ipmee} has the smallest negative baseline imbalance value at $-0.01$, whereas \texttt{shine} has the largest negative value at $-2.64$, hence the former appearing in the top of Figure \ref{Lord1to25} whereas the latter in the bottom. In Figure \ref{Lord26to50}, the top outcome is \texttt{tfl}, which has the largest positive pre-test imbalance value at $0.25$, and the bottom one is \texttt{ar}, with a baseline imbalance value at 0.

While helpful, visualisations can be misleading to the naked eye. For instance, it is rather difficult to tell when Lord's Paradox occurs, unless there is a striking difference in estimates derived from post and gain approaches in the left panels, as in \texttt{fs}, \texttt{ttsm}, and \texttt{shine} of Figure \ref{Lord1to25}, where the estimates differ remarkably in both direction and magnitude. But in either figure, there are also borderline cases where the estimates differ in sign but not much in magnitude, such as \texttt{p4cm} in Figure \ref{Lord1to25} and \texttt{efw}, \texttt{efr}, \texttt{rfr}, and \texttt{gfw} in Figure \ref{Lord26to50}. In fact, when one estimate is at or very close to the vertical zero line of the left panels, it is not straightforward to define if it is Lord's Paradox, for instance, \texttt{rp}, \texttt{cmpm}, and \texttt{uos} in Figure \ref{Lord1to25}, or \texttt{catchn}, \texttt{text}, and \texttt{ipmfm} in Figure \ref{Lord26to50}. When this is the case, it is important to refer to the statistics in either Table \ref{efs} or \ref{stats}, where \textit{pret.imb} values in the former are listed in order of signs (from negative to zero, and then to positive), whereas those in the latter are ordered from the lowest to the highest in magnitude of difference. In either case, we can choose a threshold for baseline imbalance, say, any value greater than $0.1$ in magnitude is worth a further look.

As the left panels of the two figures show, Lord's Paradox is just a special case of the reversal paradox, where the estimates differ in both direction and magnitude. But there are also occasions where the difference in magnitude might be important even though the sign remains the same, for instance, \texttt{bp}, \texttt{ttse}, \texttt{ttpm}, \texttt{ttpe}, \texttt{cmtm}, \texttt{shine} in Figure \ref{Lord1to25}, and \texttt{tfl}, \texttt{impn}, \texttt{cmpe}, \texttt{catcht}, \texttt{efm}, \texttt{ffe}, and \texttt{impl} in Figure \ref{Lord26to50}. Again, this involves human judgement. However, as in the identification of the special case, we can refer to the details in Table \ref{efs} to make cut-off decisions, for instance, any absolute difference between the difference-in-means of post-tests only, column \textit{gP}, and the different-in-means of gains, column \textit{gG}, greater than, say, 0.1 again, should be noted.

A number of issues are worth highlighting. First, whenever the treatment group has a substantially lower average attainment than that of the control group, a difference-in-means of post-tests alone tends to produce a substantially smaller estimate than that of the difference-in-means of gains, as shown in outcomes \texttt{ttse}, \texttt{ttpm}, \texttt{ttpe}, \texttt{cmtm}, \texttt{fs}, \texttt{ttsm}, and \texttt{shine} in Figure \ref{Lord1to25}. Likewise, when treated groups perform much better at base-line than control groups do, the effect estimate based on the difference-in-means of post-tests alone tends to be greater than that based on the difference-in-means of average gains, as illustrated in the following cases of Figure \ref{Lord26to50}: \texttt{tfl}, \texttt{efw}, \texttt{impn}, \texttt{efr}, \texttt{cmpe}, \texttt{catcht}, \texttt{rfr}, \texttt{gfw}, \texttt{efm}, and \texttt{catchn}. If we were to draw a line using the two point estimates from the difference-in-means of post-test and gain scores for each of the outcomes mentioned above, the slopes of the lines in Figure \ref{Lord1to25} would be very different from those for the outcomes mentioned above in Figure \ref{Lord26to50}. And for the outcomes not mentioned above in either figure, these imaginary lines would be increasingly parallel to the no-difference vertical line of zero, as their pre-test imbalance scores lean towards zero.

By and large, whenever baseline imbalance occurred, intervention groups in the EEF trials performed less well than the controlled groups did. Although there are ten negative imbalance scores ($-2.64$, $-0.41$, $-0.36$, $-0.36$, $-0.27$, $-0.21$, $-0.17$, $-0.13$, $-0.11$, $-0.11$) and ten positive ones (0.25, 0.21, 0.19, 0.19, 0.14, 0.14, 0.14, 0.14, 0.12, 0.11) that are greater than 0.10 in magnitude, the differences in magnitude are greater on the negative side. This is consistent with the interventions being targeting lower attaining pupils across the studies.

It is also worth noting that the estimates from the two simpler models do not always differ, as in, for instance, \texttt{ipmee} of Figure \ref{Lord1to25}, and \texttt{sor}, \texttt{mms}, \texttt{cbks+}, and \texttt{sar} of Figure \ref{Lord26to50}. Then comes the ``reversal'' paradox where signs remain the same while the estimates differ in magnitude. Lord's Paradox is just a special case, where controlling for a continuous baseline measure reverses the relationship between treatment status and post-test. Have examined the results from the difference-in-means models, we now turn to the estimates from the two multilevel models by focusing on the right-hand side panels of the two figures.

Let's first see what happened to the cases where the reversal paradox clearly occurred when the difference-in-means approach was employed. In Figure \ref{Lord1to25}, the most extreme outcome is \texttt{shine}, where the post and gain results differed substantially in the left panel, now they are much closer to each other in the right-hand side panel of Figure \ref{Lord1to25}. The same pattern emerges in outcomes such as \texttt{ttsm}, \texttt{fs}, \texttt{cmtm}, \texttt{ttpe}, \texttt{ttpm}, and \texttt{ttse}, where the difference-in-means models produced inconsistent, conflicting, or even contradictory estimates. The multilevel models tend to produce estimates that are more consistent than their simpler counterparts. In most cases, the post-ANCOVA multilevel model pulls the extreme estimates in the left panel of Figure \ref{Lord1to25} towards the central vertical line of zero, while the gain-ANOVA multilevel model moves less dramatically. In Figure \ref{Lord26to50}, where the outcomes are less imbalanced at baseline, the differences in point estimates or improvement made by the multilevel models are less striking. However, the models in the right-hand side panels of both figures have wider confidence intervals than those found in the left-hand side panels. This increased level of uncertainty due to the use of multilevel modelling helps avoid over-confidence in estimates from the two difference-in-means models, which generated confidence intervals that did not cross the vertical zero line for the following outcomes: \texttt{ipmee}, \texttt{ttse}, and \texttt{ttpe} in Figure \ref{Lord1to25}, and \texttt{pbcp}, \texttt{mms}, and \texttt{ipmfe} in Figure \ref{Lord26to50}.

Since point estimates in the right-hand side panels are much closer than those in the left-hand side, one might ask which multilevel model is better. As the two figures show, they produce very consistent results, except for a few cases such as \texttt{fs}, \texttt{ttsm}, \texttt{shine} in Figure \ref{Lord1to25} and \texttt{tfl} in Figure \ref{Lord26to50}, where baseline imbalance is too extreme for the two multilevel models to control for. However, it is also important to remember that using gain scores as the outcome variable involved the transformation of some pre- and post-test results which might also have had any impact on the estimate. In post-ANCOVA, the transformation was made redundant \citep[see also][p.~161]{Dimitrov2003}. While gain-ANOVA also pulls extreme estimates from the difference-in-means of gains towards the central vertical line of zero, it does so less dramatically than post-ANCOVA does to the estimates from difference-in-means of post-tests alone. This suggests either that gain-ANOVA is less powerful in constraining point estimates from the difference-in-means of gains model, or that the two models make very little difference in point estimates, though does not imply that the levels of uncertainty surrounding the point estimates are the same. As we can see from the results plotted in the right-hand side panels of the two figures, both multilevel models constrain the results from the two difference-in-means models, but gain-ANOVA pulls them more towards the centre.

While the two multilevel models produce almost identical results in most cases considered, \texttt{fs} in Figure \ref{Lord1to25} is worth a closer look. The two simpler difference-in-means models produced contradictory estimates, in fact, both point estimates and uncertainty levels were inconsistent. The two multilevel models brought the two point estimates to the same side, but one confidence interval crossed the zero line and the other did not, which might lead to different conclusions, depending on how one views statistical significance and confidence intervals. This project was highly imbalanced at pre-test, with an imbalance score of $0.36$. However, as the data in Table \ref{stats} show, its intra-cluster correlation (icc) is zero, meaning the ten schools are very similar at baseline. If we compare \texttt{fs} with \texttt{ttsm}, which is just below \texttt{fs} in Figure \ref{Lord1to25}, we can find in Table \ref{stats} that the latter is even more imbalanced at baseline, with a score of $0.41$. Lord's Paradox is identifiable in the left panel. However, the two multilevel models were able to compensate for the Paradox in the right panel. The more sophisticated models adjusted in one case but not in the other mainly because ttsm has an icc of $0.20$ (as reported in Table \ref{stats}), which indicates that the 361 schools were by far more heterogeneous than the ten in \texttt{fs} on pre-test performance. This contrast demonstrates that icc and sample sizes have a bearing on the results discussed above in relation to the estimate of effect. They also affect levels of uncertainty surrounding the point estimates from the two multilevel models.

The outcome that has the widest confidence intervals amongst all the results plotted in the right-hand side panels is \texttt{wwr} in Figure \ref{Lord26to50}. This outcome draws our attention because the uncertainty levels associated with the two simpler models are very narrow. To explain the large difference, we once again look at the statistics about the outcome in Table \ref{stats}. The icc value is $0.55$, the second highest in the study. However, the outcome with the highest icc value, \texttt{catcht} at $0.78$, does not have wider confidence intervals in the same plot. Also, \texttt{catcht} has a much smaller sample size (210 pupils) than that of \texttt{wwr} (1223 pupils). In theory, \texttt{wwr} should give the models greater confidence than catcht, particularly when \texttt{catcht} has a higher icc value and smaller sample size. But when we examine the values in the \textit{n.sch} column of Table \ref{stats}, we immediately find that the former has 16 schools whereas the latter has 54. The sample sizes in intervention and control groups are also more balanced in \texttt{catcht} (107 versus 108) than in \texttt{wwr} (658 versus 565). Upon a further look at the columns \textit{n.sch} and \textit{icc} in Table \ref{stats}, we also notice that more schools do not necessarily indicate higher icc values. However, whenever the number of schools exceeds 30, the uncertainty levels associated with the multilevel models display more consistent patterns than when the number is much smaller. This suggests that increasing the number of schools and trying to keep the sample sizes within individual schools balanced can substantially reduce the uncertainty associated with the multilevel models.

As emphasised earlier, the purpose of the study is not to check whether the effect size estimates the evaluation teams reported are ``correct'' or not. Instead, we set out to learn more about Lord's Paradox and how to resolve it if it is possible. As we have shown, the reversal paradox can occur even when only one covariate is considered. It is possible that the effect of an intervention, and the conclusion drawn from it, might no longer hold if more covariates are introduced. Given the results presented earlier, we are able to make the following suggestions for practitioners.

First, when randomisation is successful in achieving balance at baseline, or when the values in column \textit{pt.corr} of Table \ref{stats} are zero or almost zero, which means pre-test and treatment indicator are barely correlated, the two difference-in-means models produce almost identical results, as manifested in outcomes \texttt{ar}, \texttt{sar}, and \texttt{cbks+} in Figure \ref{Lord26to50}. However, when icc is relatively high, and when sample sizes in intervention and control groups are not balanced, as in \texttt{ipmee} and \texttt{ipmfe}, the results from the two simpler models may result in over-confidence in the results. In this case, multilevel models are recommended. Second, when substantial pre-test imbalance exists, as in \texttt{fs} and \texttt{ttsm}, multilevel models only reduces the inconsistency from the two simpler models when icc is as low as zero in \texttt{fs}, but they corrected Lord's Paradox when icc is as high as $0.20$ in \texttt{ttsm}. Therefore, it is important to understand when the relatively more sophisticated models can adjust successfully and how. Finally, when post and gain comparisons produce different results, we think it is better to go with the former unless research questions demand otherwise. Gain comparisons assume that the pre- and post-tests are comparable and on the same scale, which may not always be the case.

\section*{Discussion}

The empirical unravelling of Lord's Paradox demonstrates that the reversal paradox can occur in both observational and experimental studies and that Lord's Paradox is just a special case of different analytic estimates where both sign and magnitude of an effect change. In either observational or experimental studies, the relationship of interest reverses when there is substantial baseline imbalance or substantial difference in initial measure of pre-existing groups. Since randomisation does not guarantee balance in all observed covariates, even the special case of the reversal paradox may occur in any given study, not to mention unobserved covariates. As we have shown, the discrepancy in estimates from post and gain comparisons using difference-in-means can be minimised when a simple multilevel model is employed. 

Even if we are not concerned about the costs associated with baseline data collection, a simple comparison of post-test scores alone tends to underestimate an intervention effect when the intervention group has a much lower average score at baseline than that of the control group, and the difference-in-means of gains in the same case tends to overestimate the effect because of regression to the mean \citep[see also][p.~1123]{Vickers2001}. When intervention and control groups are balanced at baseline, a simple difference-in-means of either post-test or gain scores would suffice if a practitioner is only interested in the average treatment effect for the sample. However, if clusters of the sample differ substantially in their average performances and in terms of their relative sizes, the simple method is likely to produce an overly confident estimate. In such cases, a multilevel model is preferred, with post-test as the outcome when baseline and final measures are not comparable. All in all, the findings presented in this study support the conclusion that ANCOVA is ``the preferred general approach'' \citep[p.~1124]{Vickers2001}, so do those in a recent meta-analysis of RCTs published in the past 20 years about the effect of exercise on cognitive performance \citep{liusicong16}.

It is important therefore to recognise that gain and post-test comparisons to identify the impact of an educational intervention answer slightly different questions, meaning it is unrealistic to expect the same answers from them \citep{Pike2004}. A gain score comparison identifies the extent of the progress of the pupils associated with or due to the intervention, but may be biased in relation to the group with lower baseline scores where there is imbalance. A post-test comparison, controlling for pre-test (post-ANCOVA), identifies the extent of relationship or the impact of the intervention in terms of overall distribution of the pupils in the study. In some circumstances the estimates of effect from gain and post-test analyses will be the same (such as when the intervention is equally effective for all pupils and does not alter the underlying distribution). However, this is not always going to be the case, as our analysis above indicates, and may not therefore be directly comparable with each other. Based on the analysis presented earlier, we also argue that a multilevel model is to be preferred as the average treatment effect takes into account both school variation and any inequality in sample sizes in any clustered design.

However, it is also important to recognise the following limitations of the study. First, the projects selected for the exploration have varying levels of missing data. Instead of imputing the missing values, we conducted a complete case analysis, which is the headline effect size estimates in the corresponding evaluation reports. Although there are many ways to deal with missing data, and new methods are still emerging, we feel it is beyond the scope of this study. Second, like all models adopted by the evaluation teams of the projects selected for the study, we chose the four models that fit the data in varying degrees \citep[see a discussion on this topic in][]{Breiman2001}. But a model that fits the data well does not necessarily mean it is a good model, it may have a high level of explanatory power and help us estimate the impact of an intervention, but if we use it to predict outcomes for out-of-sample observations, the errors it produces might render it rather less informative \citep{ISLR15}. So, a way forward for any practitioner is to have multiple models competing on the same dataset, as we did in the study to check if the results are sensitive to the analytical methods used. But ideally, particularly when many covariates are involved, the dataset should be divided into training and test sets, and the best models are those that have the highest predictive accuracy on the test sets \citep{Donoho2015}.

Also, it is always worth bearing in mind that causal inference relies upon crucial background knowledge about an intervention, rather than universal statistical criteria \citep{Arah2008, werts1969}. The models considered and compared in the study, as are those in Lord's hypothetical case, can be all wrong when they are asked to do things they cannot deliver. They can be all correct when their assumptions are met and function in relation to their specific ``small world'' of logic. Indeed, answers to ``large world'' questions, such as the effect of an educational intervention, often depend on information that is not quantified in the data. Even if the information required is available and encoded in a model, it is still possible that the observed relationship between treatment status and the outcome of interest may be altered when another covariate is added to the model \citep[see also][p.~123]{Wainer2004}. This unwanted plausibility thus implies that statistical models are ``vulnerable to and demand critique, regardless of the precision of their estimates and apparent accuracy of their predictions'' \citep[p.~134]{McElreath2016}. But it does not suggest that statistical models are uninformative — they are helpful and will continue to play significant roles in decision science when their assumptions are theoretically grounded \cite[p.~352]{Pike2004} and used in appropriate contexts.

\subsection*{Acknowledgement}
This research was funded by a grant to Durham University from the Education Endowment Foundation. We would also like to acknowledge our thanks to those EEF evaluators, particularly Dr Ben Styles of the National Foundation for Educational Research in England and Wales, who attended a workshop on this topic and provided us with constructive feedback.

\subsection*{Conflict of interest}
The authors have no conflict of interest to disclose.

\begin{spacing}{1.2}
\bibliographystyle{apacite}
\bibliographystyle{plainnat}
\renewcommand{\bibname}{References}
\bibliography{library.bib}
\end{spacing}

\newpage
\begin{table}[!ht]
\centering
\rowcolors{1}{}{lightgray}
\resizebox{.76\textwidth}{!}{\begin{minipage}{\textwidth}
\begin{tabular}{|c|l|l|lr|}
\hline \hline 
outcome & label & full project title & design & lock\\ 
\hline
  1 & ffe & Future Foundations (English) & mst &   2 \\ 
  2 & ffm & Future Foundations (Maths) & mst &   2 \\ 
  3 & sor & Switch-on Reading & mst &   3 \\ 
  4 & gfw & Grammar for Writing & crt &   3 \\ 
  5 & rfr & Rhythm for Reading & mst &   3 \\ 
  6 & rti & Response to Intervention & crt &   1 \\ 
  7 & efr & Effective Feedback (Reading) & action &  \\ 
  8 & efw & Effective Feedback (Writing) & action &  \\ 
  9 & efm & Effective Feedback (Maths) & action &  \\ 
  10 & catchn & Catch Up Numeracy (Numeracy) & mst &   3 \\ 
  11 & catcht & Catch Up Numeracy (Time) & mst &   2 \\ 
  12 & cbks+ & Chatterbooks (Plus) & srt &   3 \\ 
  13 & cbks & Chatterbooks & srt &   3 \\ 
  14 & rp & Rapid Phonics & srt &   3 \\ 
  15 & ar & Accelerated Reader & mst &   3 \\ 
  16 & bp & Butterfly Phonics & mst &   0 \\ 
  17 & iwq & Improving Writing Quality & crt &   2 \\ 
  18 & sar & Summer Active Reading & mst &   3 \\ 
  19 & text & TextNow & mst &   3 \\ 
  20 & uos & Units of Sound & mst &   1 \\ 
  21 & ve & Vocabulary Enrichment & mst &   4 \\ 
  22 & ipmfm & Increasing Pupil Motivation (Financial - Maths) & crt &   2 \\ 
  23 & ipmfe & Increasing Pupil Motivation (Financial - English) & crt &   2 \\ 
  24 & ipmem & Increasing Pupil Motivation (Events - Maths) & crt &   2 \\ 
  25 & ipmee & Increasing Pupil Motivation (Events - English) & crt &   2 \\ 
   26 & fs & Fresh Start & mst &   3 \\ 
  27 & tfl & Talk for Literacy & mst &   4 \\ 
  28 & mms & Mathematics Mastery Secondary & crt &   4 \\ 
  29 & cmpe & Changing Mindsets (Pupil - English) & mst &   2 \\ 
  30 & cmpm & Changing Mindsets (Pupil - Maths) & mst &   2 \\ 
  31 & cmte & Changing Mindsets (Teacher - English) & crt &   3 \\ 
  32 & cmtm & Changing Mindsets (Teacher - Maths) & crt &   3 \\ 
  33 & ttpe & Tutor Trust Primary (English) & quasi &  \\ 
  34 & ttpm & Tutor Trust Primary (Maths) & quasi &  \\ 
  35 & ttsm & Tutor Trust Secondary (English) & quasi &  \\ 
  36 & ttse & Tutor Trust Secondary (Maths) & quasi &  \\ 
  37 & p4cm & Philosophy for Children (Maths) & crt &   3 \\ 
  38 & p4cr & Philosophy for Children (Reading) & crt &   3 \\ 
  39 & p4cw & Philosophy for Children (Writing) & crt &   3 \\ 
  40 & shine & SHINE in Secondaries & rdd &  \\ 
  41 & pr7 & Paired Reading (Year 7) & crt &   4 \\ 
  42 & pr9 & Paired Reading (Year 9) & crt &   4 \\ 
  43 & catchl & Catch Up Literacy & mst &   4 \\ 
  44 & pbcp & Perry Beeches Coaching Programme & mst &   3 \\ 
  45 & wwr & Word and World Reading & crt &  \\ 
  46 & quest & Quest & crt &   1 \\ 
  47 & aspm & Act, Sing, Play (Maths) & mst &   4 \\ 
  48 & aspl & Act, Sing, Play (Literacy) & mst &   4 \\ 
  49 & impn & Improving Numeracy and Literacy (Numeracy) & crt &   5 \\ 
  50 & impl & Improving Numeracy and Literacy (Literacy) & crt &   5 \\ 
\hline \hline
\end{tabular}
\centering
\caption{{\bf 50 intervention outcomes from 34 EEF projects.} \texttt{outcome} lists 50 intervention outcomes, each having an effect size estimation. \texttt{label} is outcome abbreviations. \texttt{full project title} is the title used for each project funded by the EEF. Note that some projects have multiple interventions, hence multiple outcomes. \texttt{design} indicates the type of research design for a project, where \texttt{mst}, \texttt{crt}, and \texttt{srt} are multi-site, cluster, and simple randomised trials respectively (for the definitions of the three designs, please see \cite[p.~4]{zmxaksh2016a}. \texttt{action} is action research, \texttt{quasi} refers to well-matched quasi-experiments. \texttt{rdd} is a regression discontinuity design. \texttt{lock} signifies the quality of design and implementation for causal inference, the higher the value, the more valid the inference would be. Some cells are empty where a padlock was not assigned because the study was a pilot or has only weak causal inference.}
\label{fullname}
\end{minipage} }

\end{table}

\newpage
\begin{table}[!ht]
\rowcolors{1}{}{lightgray}
\resizebox{.839\textwidth}{!}{\begin{minipage}{1\textwidth}
\centering
\begin{tabular}{|l|ccc|ccc|ccc|ccc|c|}
  \hline \hline
label & gP & gP.lb & gP.ub & gG & gG.lb & gG.ub & ttP & ttP.lb & ttP.ub & ttG & ttG.lb & ttG.ub & pret.imb \\ 
  \hline
shine & -1.66 & -1.85 & -1.46 & -0.02 & -0.19 & 0.14 & 0.14 & -0.16 & 0.45 & -0.02 & -0.31 & 0.27 & -2.64 \\ 
  ttsm & -0.23 & -0.30 & -0.16 & 0.27 & 0.20 & 0.34 & 0.08 & -0.22 & 0.39 & 0.15 & -0.14 & 0.43 & -0.41 \\ 
  fs & -0.20 & -0.40 & -0.01 & 0.24 & 0.05 & 0.43 & 0.07 & -0.23 & 0.37 & 0.24 & 0.05 & 0.43 & -0.36 \\ 
  cmtm & -0.28 & -0.42 & -0.15 & -0.07 & -0.20 & 0.07 & -0.04 & -0.29 & 0.21 & -0.09 & -0.33 & 0.15 & -0.36 \\ 
  ttpe & -0.53 & -0.76 & -0.30 & -0.31 & -0.53 & -0.08 & -0.15 & -0.70 & 0.40 & -0.11 & -0.52 & 0.30 & -0.27 \\ 
  ttpm & -0.40 & -0.67 & -0.14 & -0.25 & -0.51 & 0.02 & -0.21 & -0.68 & 0.26 & -0.12 & -0.55 & 0.31 & -0.21 \\ 
  ttse & -0.31 & -0.41 & -0.22 & -0.18 & -0.28 & -0.08 & -0.12 & -0.59 & 0.35 & -0.08 & -0.51 & 0.34 & -0.17 \\ 
  p4cm & -0.04 & -0.15 & 0.06 & 0.10 & -0.01 & 0.21 & -0.01 & -0.29 & 0.26 & 0.04 & -0.23 & 0.31 & -0.13 \\ 
  uos & -0.09 & -0.28 & 0.10 & 0.01 & -0.18 & 0.20 & -0.04 & -0.29 & 0.21 & 0.00 & -0.25 & 0.26 & -0.11 \\ 
  cmpm & -0.01 & -0.31 & 0.28 & 0.12 & -0.17 & 0.42 & 0.15 & -0.20 & 0.50 & 0.12 & -0.20 & 0.45 & -0.11 \\ 
  rp & -0.09 & -0.39 & 0.20 & 0.01 & -0.29 & 0.30 & -0.05 & -0.34 & 0.25 & 0.01 & -0.29 & 0.30 & -0.10 \\ 
  cmte & -0.20 & -0.33 & -0.06 & -0.12 & -0.25 & 0.01 & -0.19 & -0.39 & 0.01 & -0.13 & -0.37 & 0.10 & -0.10 \\ 
  p4cr & 0.02 & -0.09 & 0.13 & 0.11 & 0.01 & 0.22 & 0.04 & -0.18 & 0.27 & 0.07 & -0.15 & 0.30 & -0.09 \\ 
  pr9 & -0.12 & -0.23 & -0.01 & -0.09 & -0.20 & 0.02 & -0.11 & -0.40 & 0.18 & -0.10 & -0.37 & 0.16 & -0.08 \\ 
  catchl & 0.08 & -0.09 & 0.25 & 0.17 & -0.01 & 0.34 & 0.14 & -0.14 & 0.43 & 0.15 & -0.17 & 0.47 & -0.07 \\ 
  aspm & -0.05 & -0.20 & 0.09 & -0.01 & -0.15 & 0.14 & -0.01 & -0.34 & 0.32 & 0.00 & -0.32 & 0.32 & -0.06 \\ 
  iwq & 0.59 & 0.35 & 0.84 & 0.59 & 0.35 & 0.84 & 0.67 & 0.07 & 1.28 & 0.54 & -0.02 & 1.10 & -0.05 \\ 
  p4cw & -0.07 & -0.18 & 0.03 & -0.03 & -0.14 & 0.08 & -0.14 & -0.41 & 0.12 & -0.11 & -0.37 & 0.14 & -0.05 \\ 
  pr7 & -0.07 & -0.18 & 0.04 & -0.04 & -0.15 & 0.07 & -0.04 & -0.23 & 0.16 & -0.03 & -0.19 & 0.13 & -0.05 \\ 
  cbks & -0.07 & -0.30 & 0.15 & -0.06 & -0.28 & 0.17 & -0.08 & -0.40 & 0.24 & -0.06 & -0.35 & 0.24 & -0.03 \\ 
  ve & 0.04 & -0.13 & 0.20 & 0.09 & -0.08 & 0.25 & 0.07 & -0.18 & 0.32 & 0.08 & -0.17 & 0.34 & -0.03 \\ 
  quest & -0.05 & -0.14 & 0.03 & -0.03 & -0.12 & 0.05 & -0.01 & -0.27 & 0.25 & 0.02 & -0.26 & 0.31 & -0.02 \\ 
  aspl & 0.01 & -0.14 & 0.15 & 0.04 & -0.11 & 0.19 & 0.03 & -0.16 & 0.22 & 0.04 & -0.13 & 0.21 & -0.02 \\ 
  bp & 0.32 & 0.09 & 0.55 & 0.45 & 0.22 & 0.68 & 0.47 & 0.06 & 0.87 & 0.44 & 0.03 & 0.85 & -0.01 \\ 
  ipmee & -0.07 & -0.12 & -0.02 & -0.07 & -0.13 & -0.02 & 0.00 & -0.26 & 0.25 & -0.01 & -0.23 & 0.21 & -0.01 \\ 
  ar & 0.25 & 0.03 & 0.47 & 0.28 & 0.06 & 0.50 & 0.32 & -0.02 & 0.66 & 0.28 & 0.06 & 0.50 & 0.00 \\ 
  sar & 0.12 & -0.17 & 0.41 & 0.11 & -0.18 & 0.40 & 0.14 & -0.20 & 0.47 & 0.12 & -0.21 & 0.45 & 0.00 \\ 
  cbks+ & 0.03 & -0.19 & 0.26 & 0.03 & -0.20 & 0.25 & 0.03 & -0.31 & 0.37 & 0.03 & -0.30 & 0.37 & 0.01 \\ 
  ipmfe & -0.05 & -0.10 & 0.00 & -0.07 & -0.12 & -0.02 & -0.02 & -0.26 & 0.22 & -0.03 & -0.23 & 0.17 & 0.01 \\ 
  mms & 0.08 & 0.03 & 0.13 & 0.07 & 0.02 & 0.12 & 0.08 & -0.10 & 0.26 & 0.06 & -0.11 & 0.23 & 0.02 \\ 
  sor & 0.24 & 0.01 & 0.46 & 0.24 & 0.02 & 0.47 & 0.30 & 0.03 & 0.56 & 0.26 & 0.01 & 0.51 & 0.05 \\ 
  pbcp & 0.36 & 0.13 & 0.59 & 0.39 & 0.16 & 0.63 & 0.43 & -0.08 & 0.94 & 0.40 & -0.16 & 0.96 & 0.05 \\ 
  ffm & 0.01 & -0.21 & 0.24 & -0.05 & -0.27 & 0.18 & -0.04 & -0.35 & 0.27 & -0.09 & -0.44 & 0.26 & 0.06 \\ 
  rti & 0.17 & -0.03 & 0.37 & 0.15 & -0.05 & 0.35 & 0.14 & -0.22 & 0.49 & 0.14 & -0.15 & 0.42 & 0.06 \\ 
  ipmfm & 0.07 & 0.02 & 0.12 & 0.01 & -0.04 & 0.06 & 0.11 & -0.12 & 0.34 & 0.06 & -0.14 & 0.27 & 0.06 \\ 
  impl & -0.06 & -0.17 & 0.04 & -0.22 & -0.32 & -0.11 & -0.08 & -0.35 & 0.20 & -0.10 & -0.37 & 0.16 & 0.07 \\ 
  ffe & 0.19 & -0.03 & 0.41 & 0.12 & -0.10 & 0.34 & 0.14 & -0.29 & 0.57 & 0.09 & -0.35 & 0.53 & 0.08 \\ 
  text & -0.01 & -0.21 & 0.19 & -0.09 & -0.29 & 0.11 & -0.07 & -0.33 & 0.19 & -0.08 & -0.36 & 0.20 & 0.08 \\ 
  ipmem & 0.06 & 0.01 & 0.11 & -0.03 & -0.08 & 0.02 & 0.04 & -0.21 & 0.28 & -0.01 & -0.21 & 0.20 & 0.08 \\ 
  wwr & 0.02 & -0.10 & 0.13 & -0.04 & -0.15 & 0.08 & 0.04 & -0.79 & 0.87 & 0.04 & -0.79 & 0.87 & 0.08 \\ 
  catchn & 0.00 & -0.27 & 0.26 & -0.11 & -0.38 & 0.16 & -0.03 & -0.37 & 0.31 & -0.11 & -0.43 & 0.21 & 0.11 \\ 
  efm & 0.15 & 0.07 & 0.22 & 0.05 & -0.02 & 0.13 & -0.05 & -0.58 & 0.48 & -0.09 & -0.67 & 0.48 & 0.12 \\ 
  gfw & 0.11 & 0.00 & 0.21 & -0.03 & -0.13 & 0.08 & 0.08 & -0.14 & 0.31 & 0.00 & -0.21 & 0.21 & 0.14 \\ 
  rfr & 0.12 & -0.09 & 0.32 & -0.02 & -0.23 & 0.18 & 0.03 & -0.51 & 0.57 & -0.05 & -0.51 & 0.42 & 0.14 \\ 
  catcht & 0.19 & -0.08 & 0.46 & 0.11 & -0.16 & 0.38 & 0.12 & -0.24 & 0.48 & 0.02 & -0.34 & 0.39 & 0.14 \\ 
  cmpe & 0.26 & -0.03 & 0.56 & 0.15 & -0.14 & 0.45 & 0.24 & -0.06 & 0.53 & 0.15 & -0.25 & 0.54 & 0.14 \\ 
  efr & 0.18 & 0.11 & 0.26 & -0.04 & -0.11 & 0.04 & 0.00 & -0.28 & 0.29 & -0.09 & -0.49 & 0.31 & 0.19 \\ 
  impn & 0.32 & 0.21 & 0.43 & 0.21 & 0.10 & 0.32 & 0.30 & 0.02 & 0.58 & 0.24 & -0.04 & 0.53 & 0.19 \\ 
  efw & 0.18 & 0.11 & 0.26 & -0.05 & -0.12 & 0.03 & -0.05 & -0.39 & 0.29 & -0.10 & -0.50 & 0.29 & 0.21 \\ 
  tfl & 0.35 & 0.08 & 0.62 & 0.11 & -0.15 & 0.38 & 0.25 & -0.08 & 0.57 & 0.11 & -0.17 & 0.40 & 0.25 \\ 
   \hline \hline
\end{tabular}
\end{minipage} }
\captionsetup{width=1\textwidth}
\caption{ {\bf Effect size estimates listed in ascending order of pre-test imbalance scores.} \texttt{gP} and \texttt{gG} contain estimates from the difference-in-means approach, with the former using post-test only and the latter using gain score as the outcome. \texttt{ttP} and \texttt{ttG} present results from two MLMs, post-ANCOVA and gain-ANOVA respectively. \texttt{lb} and \texttt{ub} refer to lower and upper bounds of 95\% confidence intervals.}
\label{efs}
\end{table}

\newpage
\begin{table}[!ht]
\centering
\rowcolors{1}{}{lightgray}
\resizebox{.83\textwidth}{!}{\begin{minipage}{\textwidth}
\begin{tabular}{|l|cccc|ccc|c|}
  \hline \hline
label & n & n.t & n.c & n.sch & icc & pt.corr & pp.corr & pret.imb \\ 
  \hline
ar & 326 & 167 & 159 &   4 & 0.00 & 0.00 & 0.62 & 0.00 \\ 
  sar & 182 &  93 &  89 &  48 & 0.13 & 0.00 & 0.43 & 0.00 \\ 
  cbks+ & 303 & 154 & 149 &  12 & 0.09 & 0.00 & 0.63 & 0.01 \\ 
  bp & 302 & 159 & 143 &   6 & 0.09 & -0.01 & 0.71 & -0.01 \\ 
  ipmfe & 7182 & 2298 & 4884 &  48 & 0.09 & 0.01 & 0.66 & 0.01 \\ 
  ipmee & 6950 & 2066 & 4884 &  48 & 0.11 & 0.00 & 0.65 & -0.01 \\ 
  mms & 5830 & 3197 & 2633 &  44 & 0.06 & 0.01 & 0.67 & 0.02 \\ 
  quest & 2090 & 938 & 1152 &  19 & 0.08 & -0.01 & 0.71 & -0.02 \\ 
  aspl & 814 & 541 & 273 &  19 & 0.02 & -0.01 & 0.75 & -0.02 \\ 
  cbks & 311 & 162 & 149 &  12 & 0.05 & -0.01 & 0.65 & -0.03 \\ 
  ve & 570 & 282 & 288 &  12 & 0.05 & -0.02 & 0.68 & -0.03 \\ 
  sor & 308 & 155 & 153 &  19 & 0.03 & 0.02 & 0.65 & 0.05 \\ 
  iwq & 265 & 144 & 121 &  22 & 0.24 & -0.02 & 0.41 & -0.05 \\ 
  p4cw & 1326 & 637 & 689 &  45 & 0.14 & -0.03 & 0.68 & -0.05 \\ 
  pr7 & 1309 & 627 & 682 &  10 & 0.02 & -0.02 & 0.81 & -0.05 \\ 
  pbcp & 291 & 149 & 142 &   4 & 0.11 & 0.03 & 0.65 & 0.05 \\ 
  ffm & 303 & 162 & 141 &  33 & 0.22 & 0.03 & 0.52 & 0.06 \\ 
  rti & 373 & 178 & 195 &  48 & 0.10 & 0.03 & 0.69 & 0.06 \\ 
  ipmfm & 7198 & 2311 & 4887 &  48 & 0.10 & 0.03 & 0.69 & 0.06 \\ 
  aspm & 816 & 542 & 274 &  19 & 0.18 & -0.03 & 0.72 & -0.06 \\ 
  catchl & 528 & 267 & 261 &  15 & 0.14 & -0.03 & 0.60 & -0.07 \\ 
  impl & 1427 & 577 & 850 &  38 & 0.12 & 0.03 & 0.81 & 0.07 \\ 
  ffe & 310 & 167 & 143 &  33 & 0.45 & 0.04 & 0.57 & 0.08 \\ 
  text & 391 & 199 & 192 &  54 & 0.28 & 0.04 & 0.53 & 0.08 \\ 
  ipmem & 6961 & 2074 & 4887 &  48 & 0.09 & 0.04 & 0.69 & 0.08 \\ 
  pr9 & 1276 & 620 & 656 &  10 & 0.07 & -0.04 & 0.78 & -0.08 \\ 
  wwr & 1223 & 658 & 565 &  16 & 0.55 & 0.04 & 0.62 & 0.08 \\ 
  p4cr & 1325 & 637 & 688 &  45 & 0.10 & -0.04 & 0.53 & -0.09 \\ 
  rp & 178 &  86 &  92 &  21 & 0.00 & -0.05 & 0.59 & -0.10 \\ 
  cmte & 885 & 359 & 526 &  24 & 0.05 & -0.05 & 0.79 & -0.10 \\ 
  catchn & 215 & 107 & 108 &  54 & 0.41 & 0.06 & 0.44 & 0.11 \\ 
  uos & 427 & 225 & 202 &  33 & 0.12 & -0.05 & 0.71 & -0.11 \\ 
  cmpm & 174 &  88 &  86 &   5 & 0.01 & -0.06 & 0.84 & -0.11 \\ 
  efm & 2851 & 1677 & 1174 &  14 & 0.21 & 0.06 & 0.88 & 0.12 \\ 
  p4cm & 1326 & 637 & 689 &  45 & 0.16 & -0.06 & 0.65 & -0.13 \\ 
  gfw & 1367 & 667 & 700 &  50 & 0.19 & 0.07 & 0.30 & 0.14 \\ 
  rfr & 355 & 175 & 180 &   6 & 0.10 & 0.07 & 0.62 & 0.14 \\ 
  catcht & 210 & 102 & 108 &  54 & 0.78 & 0.07 & 0.59 & 0.14 \\ 
  cmpe & 178 &  89 &  89 &   5 & 0.04 & 0.07 & 0.75 & 0.14 \\ 
  ttse & 63379 & 409 & 62970 & 281 & 0.28 & -0.01 & 0.68 & -0.17 \\ 
  efr & 2849 & 1676 & 1173 &  14 & 0.10 & 0.09 & 0.88 & 0.19 \\ 
  impn & 1365 & 517 & 848 &  36 & 0.13 & 0.09 & 0.74 & 0.19 \\ 
  efw & 2826 & 1649 & 1177 &  14 & 0.10 & 0.10 & 0.88 & 0.21 \\ 
  ttpm & 724 &  59 & 665 &  11 & 0.09 & -0.06 & 0.66 & -0.21 \\ 
  tfl & 213 & 106 & 107 &   3 & 0.00 & 0.13 & 0.61 & 0.25 \\ 
  ttpe & 786 &  83 & 703 &   9 & 0.07 & -0.08 & 0.66 & -0.27 \\ 
  fs & 419 & 215 & 204 &  10 & 0.00 & -0.18 & 0.72 & -0.36 \\ 
  cmtm & 896 & 358 & 538 &  24 & 0.05 & -0.17 & 0.77 & -0.36 \\ 
  ttsm & 101772 & 781 & 100991 & 361 & 0.20 & -0.04 & 0.79 & -0.41 \\ 
  shine & 549 & 283 & 266 &   4 & 0.03 & -0.80 & 0.81 & -2.64 \\ 
   \hline \hline
\end{tabular}
\centering
\captionsetup{width=.82\textwidth}
\caption{{\bf Summary statistics for the 50 outcomes examined, listed in ascending order of absolute pre-test imbalance values.} \texttt{n} is the sample sizes used for this study, where \texttt{n.t} and \texttt{n.c} are sample sizes for treatment and control groups. \texttt{n.sch} is the number of schools for each study. \texttt{icc} is intra-cluster correlation. \texttt{pret.imb} is pre-test imbalance. \texttt{pt.corr} and \texttt{pp.corr} are pre-test versus treatment status and pre-test versus post-test correlations, respectively.}
\label{stats}
\end{minipage} }
\end{table}

\newpage

\newpage

\end{document}